\documentclass[a4paper,aps,prd,10pt,preprintnumbers,showpacs,twocolumn,superscriptaddress,nofootinbib,amsmath,amssymb]{revtex4-1}
\usepackage{cases}
\usepackage{xcolor}
\usepackage{graphicx}
\usepackage[utf8]{inputenc}
\usepackage[T1]{fontenc}
\usepackage{cmap}
\usepackage[normalem]{ulem} 
\def\imo{i}

\def\K{{\cal K}}

\begin{document}
\title{Ringing of the regular black-hole/wormhole transition}
\author{M. S. Churilova}\email{wwrttye@gmail.com}
\affiliation{Research Centre for Theoretical Physics and Astrophysics, Institute of Physics, Silesian University in Opava, CZ-746 01 Opava, Czech Republic}
\author{Z. Stuchlík}\email{zdenek.stuchlik@fpf.slu.cz}
\affiliation{Research Centre for Theoretical Physics and Astrophysics, Institute of Physics, Silesian University in Opava, CZ-746 01 Opava, Czech Republic}
\begin{abstract}
A simple one-parameter generalization of the Schwarzschild spacetime was recently suggested by A. Simpson and M. Visser [JCAP 1902, 042 (2019)] as a toy model describing the regular black hole and traversable wormhole states separated by the border (one-way wormhole) state. We study quasinormal modes of all the three states and show that the black-hole/wormhole transition is characterized by echoes, while the remnant of the black hole state is kept in the time-domain profile of the wormhole perturbation at the initial stage of the exponential fall off. Calculations of quasinormal modes using the WKB method with Pad\'{e} expansion and the time-domain integration are in good agreement. An analytical formula governing quasinormal modes in the eikonal regime is given.
\end{abstract}
\pacs{04.50.Kd,04.70.Bw,04.30.-w,04.80.Cc}
\maketitle

\section{Introduction}

The current observations in the gravitational and electromagnetic spectra, although favorable to black holes \cite{Abbott:2016blz,TheLIGOScientific:2016src,Goddi:2017pfy,Akiyama:2019cqa}, do not exclude more exotic objects such as exotic stars \cite{Camilo:2018goy,Konoplya:2019nzp} and wormholes  \cite{Damour:2007ap,Cardoso:2016rao,Konoplya:2016pmh}.

The wormholes can be, theoretically, designed in such a way that their behavior is indistinguishable from black holes, particularly in gravitational radiation and astrophysical optical effects
\cite{Damour:2007ap,Cardoso:2016rao,Konoplya:2016pmh}.  However, a symmetric (with respect to the throat) wormhole can be distinguished from a black hole somewhat easier as it cannot mimic simultaneously the black hole ringing at a few dominant multipoles \cite{Konoplya:2016hmd}. This is caused by a qualitative difference between the symmetric (with respect to the throat) wormhole and black hole effective potentials. This is why a symmetric (with respect to the throat) wormhole can be designed to mimic the black hole ringing only at one multipole.

One possible fingerprint of a compact object is the spectrum of its quasinormal modes which depends on the parameters of the geometry \cite{Konoplya:2011qq,Kokkotas:1999bd,Berti:2009kk}. Quasinormal modes of black holes, recently observed in the mergers of two black holes \cite{Abbott:2016blz,TheLIGOScientific:2016src}, are regarded as black holes' oscillation proper frequencies under the requirement of purely outgoing wave at infinity and purely incoming wave at the event horizon. This means absence of incoming waves from both plus- and minus- infinity in terms of the tortoise coordinate.

Quasinormal modes of the wormholes (connecting two asymptotic regions) obey the same boundary conditions in terms of the tortoise coordinate as those for black holes and they were studied in various contexts in
\cite{Konoplya:2005et,Konoplya:2016hmd,Konoplya:2010kv,Roy:2019yrr,Ovgun:2019yor,Bronnikov:2012ch,Kim:2018ang,Oliveira:2018oha,Blazquez-Salcedo:2018ipc,Cuyubamba:2018jdl,Konoplya:2018ala,Aneesh:2018hlp,Bronnikov:2019sbx,Churilova:2019qph}

An interesting model interpolating between the traversable wormhole and regular black hole states was suggested in \cite{Simpson:2018tsi} and extended to the time-dependent case in \cite{Simpson:2019cer}. Here we would like to see how the regular black hole/wormhole transition via the black bounce can look like from the point of view of quasinormal ringing.

As the spacetime suggested in \cite{Simpson:2018tsi} is a one-parameter family of static geometries, the black hole/wormhole transition means shifting through this family while the parameter changes continuously. Supposing that (as a result of some physical process) the parameter continuous changes are sufficiently slow, we obtain an effective static geometry \cite{Abdalla:2006vb}.

We show that the black-hole/wormhole transition is characterized by echoes and that the continuous change of the parameter implies three stages of the ringing:
\begin{enumerate}
\item the usual ringing of the regular black hole,
\item remnants of the first stage of the black-hole ringing, followed by a series of echoes, describing the wormhole near the transition, which go over into
\item the distinctive wormhole's mode.
\end{enumerate}
 We perform the calculations of quasinormal modes using the WKB method with Pad\'{e} expansion or the time-domain integration, and demonstrate that the results coincide with high precision. We also obtain an analytical formula for quasinormal modes in the eikonal regime.

\begin{figure*}
\resizebox{\linewidth}{!}{\includegraphics*{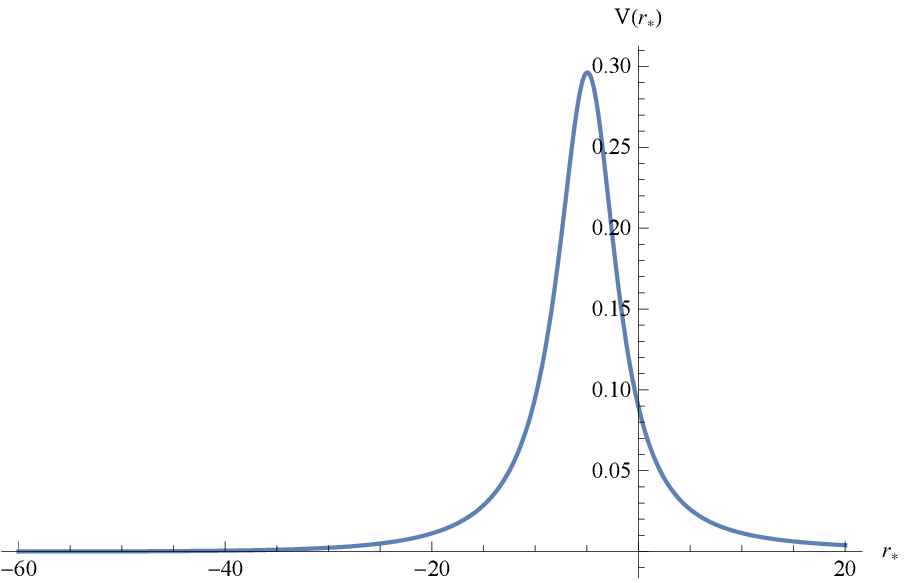}\includegraphics*{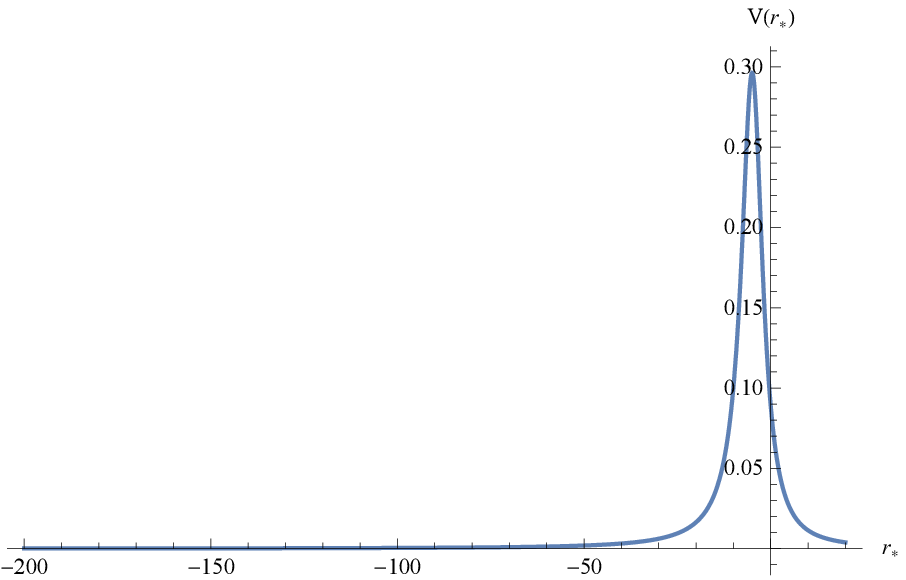}}
\caption{Effective potentials in the cases $a=0.99$, $a=1.00$ (from left to right, with $M=0.5$) for the electromagnetic field ($s=1$) and $\ell=1$.}
\label{pot0}
\end{figure*}

The paper is organized as follows. In Sec II we introduce the metric under consideration, master equations for scalar and electromagnetic perturbations, and the related effective potentials. We also outline the methods we use: the WKB method with Pad\'{e} approximants, and the time-domain integration. In Sec. III we present quasinormal modes for scalar and electromagnetic fields, demonstrate the effects of echoes in the black hole/wormhole transition and describe three observed types of ringing behaviour. We list the successive time-domain profiles, which helps to evaluate these effects qualitatively. The reliability of the obtained results is confirmed by agreement of the calculations in frequency and time domains. For the eikonal regime we find an analytical formula for quasinormal modes. In Conclusions we summarize obtained results and mention some open problems.

\begin{figure*}
\resizebox{\linewidth}{!}{\includegraphics*{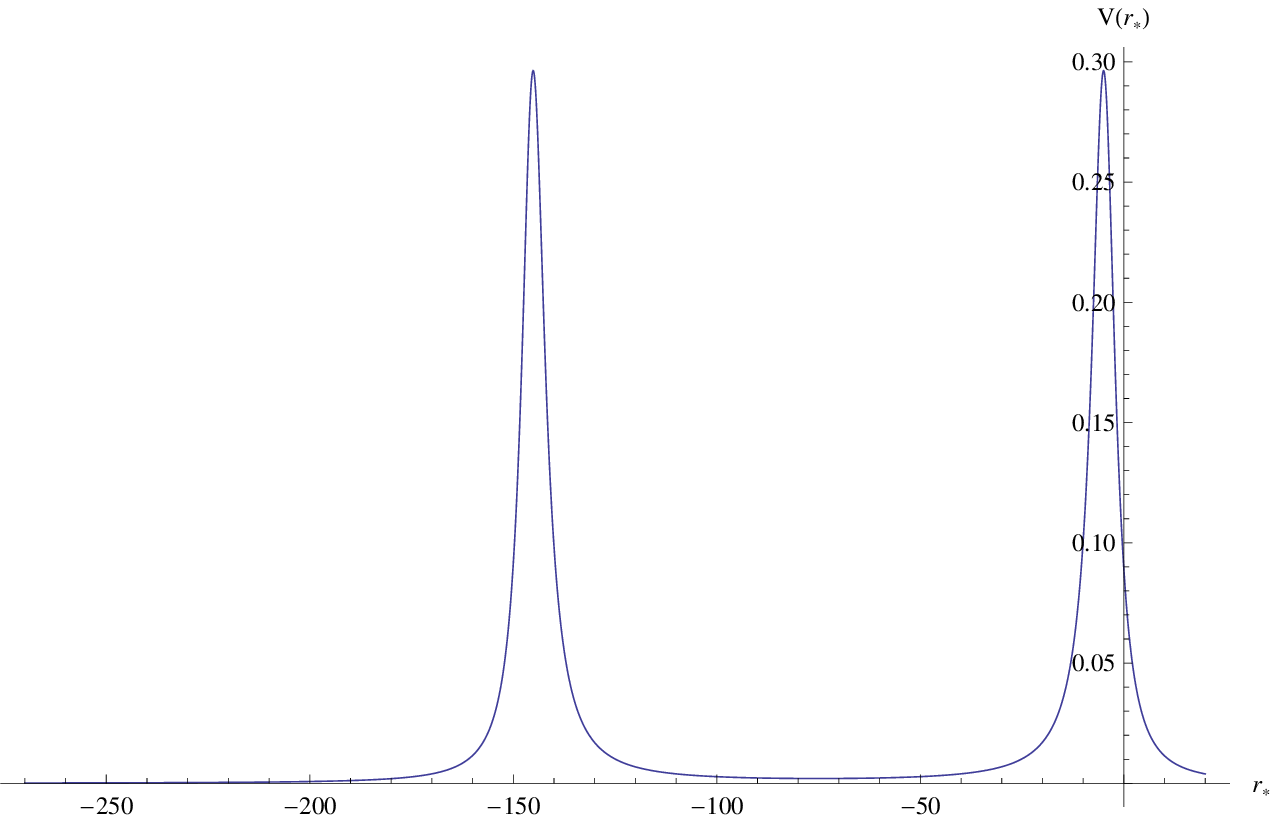}\includegraphics*{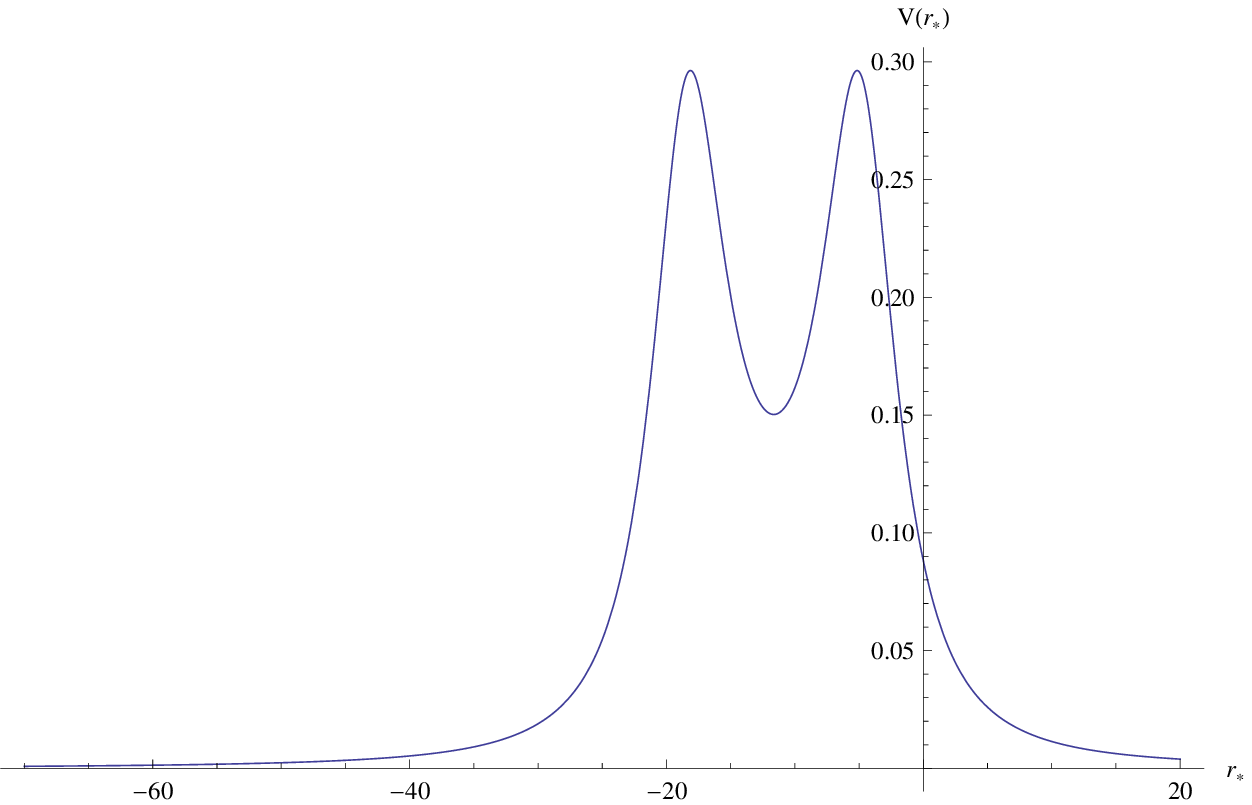}}
\caption{Effective potentials near the transition zone in the cases $a=1.001$, $a=1.100$ (from left to right, with $M=0.5$) for the electromagnetic field ($s=1$) and $\ell=1$.}
\label{pot1}
\end{figure*}

\section{The methods}

\subsection{The wave equation}

We consider scalar and electromagnetic perturbations of the static spherically symmetric spacetime described by the metric suggested in \cite{Simpson:2018tsi}:
$$
d s^2 = -\left(1- \frac{2M}{\sqrt{r^2+a^2}}\right) d t^2 + \frac{1}{1- \frac{2M}{\sqrt{r^2+a^2}}} d r^2+
$$
\begin{equation}\label{metric}
 +\left(r^2+a^2\right)\left(\sin^2\theta d\phi^2+d\theta^2\right)\,,
\end{equation}
which interpolates between black hole and wormhole states. For different values of the non-negative parameter $a$ this metric specifies correspondingly:
\begin{itemize}
\item  the Schwarzschild black hole for $a=0$;
\item  a regular black hole for $a \in (0,2M)$;
\item  a one-way wormhole with an extremal null throat for $a=2M$;
\item  the Morris-Thorne traversable wormhole for $a>2M$.
\end{itemize}

 The general covariant equation governing a massless scalar field takes the form
\begin{equation}\label{KGg}
\frac{1}{\sqrt{-g}}\partial_\mu \left(\sqrt{-g}g^{\mu \nu}\partial_\nu\Phi\right)=0
\end{equation}
and for an electromagnetic field it has the form
\begin{equation}\label{EmagEq}
\frac{1}{\sqrt{-g}}\partial_\mu \left(F_{\rho\sigma}g^{\rho \nu}g^{\sigma \mu}\sqrt{-g}\right)=0\,,
\end{equation}
where $F_{\rho\sigma}=\partial_\rho A^\sigma-\partial_\sigma A^\rho$ and $A_\mu$ is a vector potential.
After separation of the variables Eqs. (\ref{KGg}) and (\ref{EmagEq}) take the following general wave-like form
\begin{equation}\label{wave-equation}
\frac{d^2\Psi_s}{dr_*^2}+\left(\omega^2-V(r)\right)\Psi_s=0,
\end{equation}
where $s=0$ corresponds to scalar field and $s=1$ to electromagnetic field, the "tortoise coordinate" $r_*$ is defined by the relation
\begin{equation}
dr_*=\frac{dr}{1- \frac{2M}{\sqrt{r^2+a^2}}}
\end{equation}
and the effective potential reads
$$
V(r)=\left(1- \frac{2M}{\sqrt{r^2+a^2}}\right)\left(\frac{\ell\left(\ell+1\right)}{r^2+a^2}+\right.
$$
\begin{equation}\label{potentialScalar}
\left.+\left(1-s\right)
\left(\frac{2M\left(r^2-a^2\right)+a^2\sqrt{r^2+a^2}}{\left(r^2+a^2\right)^{5/2}}\right)\right).
\end{equation}

While the effective potential for the regular black-hole case, $a < 2M$, has the form of a potential barrier decaying at both infinities (see Fig. \ref{pot0}), after the transition to the wormhole state, a second peak appears far from the initial black-hole peak and then, as $a$ is increased (see Fig. \ref{pot1}) this second peak approaches the first one merging at some larger $a$. Such a behavior is qualitatively similar to the case of Damour-Solodukhin wormhole \cite{Damour:2007ap} and, as it was shown in \cite{Cardoso:2016rao} this second peak causes far from the surface (throat) of the compact object modification of a signal at late times called echoes.

\subsection{The WKB method}

In the frequency domain we use the WKB method \cite{Schutz:1985zz,Iyer:1986np,Konoplya:2003ii,Matyjasek:2017psv,Konoplya:2019hlu} and its extension to the form of the Pad\'{e} approximants.

The WKB formula developed in \cite{MashhoonBlome,Schutz:1985zz,Iyer:1986np,Konoplya:2003ii,Matyjasek:2017psv,Konoplya:2019hlu} is appropriate for finding quasinormal modes, if the effective potential has the form of the potential barrier illustrated in Fig. \ref{pot0}, which provides us with two turning points in the wave equation (\ref{wave-equation}). The WKB method is based on the expansion of the wave function into the WKB series in the asymptotic regions, matched with the Taylor expansion near the peak of the effective potential through these two turning points. As the WKB method is discussed in a numerous literature and recently has been surveyed in \cite{Konoplya:2019hlu}, we will briefly state its fundamentals.

To find quasinormal modes we use the higher-order WKB formula \cite{Konoplya:2019hlu}:
\begin{eqnarray}\label{WKBformula-spherical}
\omega^2&=&V_0+A_2(\K^2)+A_4(\K^2)+A_6(\K^2)+\ldots\\\nonumber&-&\imo \K\sqrt{-2V_2}\left(1+A_3(\K^2)+A_5(\K^2)+A_7(\K^2)\ldots\right),
\end{eqnarray}
where $\K$ takes half-integer values. The corrections $A_k(\K^2)$ of order $k$ to the first-order formula are polynomials of $\K^2$ with rational coefficients. They depend on the values $V_2$, $V_3$, ... of higher derivatives of the potential $V(r)$ in its maximum and do not depend on the maximum $V_0$ itself, which implies that the right-hand side of the formula (\ref{WKBformula-spherical}) does not depend on $\omega$. The parameter of expansion
$$
\epsilon=\frac{-2\left(\omega^2-V_0\right)}{V_2}
$$
being small defines also the region of validity of the WKB approximation.

 To increase accuracy of the WKB formula, we follow Matyjasek and Opala \cite{Matyjasek:2017psv} and use Padé approximants. For the order $k$ of the WKB formula (\ref{WKBformula-spherical}) we define a polynomial $P_k(\epsilon)$ as
\begin{eqnarray}\nonumber
  P_k(\epsilon)&=&V_0+A_2(\K^2)\epsilon^2+A_4(\K^2)\epsilon^4+A_6(\K^2)\epsilon^6+\ldots\\&-&\imo \K\sqrt{-2V_2}\left(\epsilon+A_3(\K^2)\epsilon^3+A_5(\K^2)\epsilon^5\ldots\right),\label{WKBpoly}
\end{eqnarray}
where the formal parameter $\epsilon$ keeps track of orders in the WKB approximation. Then the squared frequency is obtained for $\epsilon=1$:
$$\omega^2=P_k(1).$$

For the polynomial $P_k(\epsilon)$ we consider a family of the rational functions
\begin{equation}\label{WKBPade}
P_{\tilde{n}/\tilde{m}}(\epsilon)=\frac{Q_0+Q_1\epsilon+\ldots+Q_{\tilde{n}}\epsilon^{\tilde{n}}}{R_0+R_1\epsilon+\ldots+R_{\tilde{m}}\epsilon^{\tilde{m}}},
\end{equation}
called Padé approximants, with $\tilde{n}+\tilde{m}=k$, such that near $\epsilon=0$
$$P_{\tilde{n}/\tilde{m}}(\epsilon)-P_k(\epsilon)={\cal O}\left(\epsilon^{k+1}\right).$$

Pad\'{e} approximation essentially (up to 1 order) increases the accuracy of the WKB method, allowing to make a guess about asymptotic behavior of the WKB series.  More information on the Pad\'{e} approximants can be found in \cite{Wuytack}.

Generally, for finding fundamental mode ($n=0$), the Padé approximants with $\tilde{n}\approx\tilde{m}$ give the best approximation. In \cite{Matyjasek:2017psv}, $P_{6/6}(1)$ and $P_{6/7}(1)$ were compared to the 6th-order WKB formula $P_{6/0}(1)$. In \cite{Konoplya:2019hlu} it has been observed that usually even $P_{3/3}(1)$, i.~e. the Padé approximation of the 6th-order, gives a more accurate value for the squared frequency than $P_{6/0}(1)$.
We use this observation to find appropriate Padé partition in our case. The corresponding automatic code in \emph{Mathematica}\circledR $\;$ is in open access \cite{WKB-code}.

\begin{figure*}
\resizebox{\linewidth}{!}{\includegraphics*{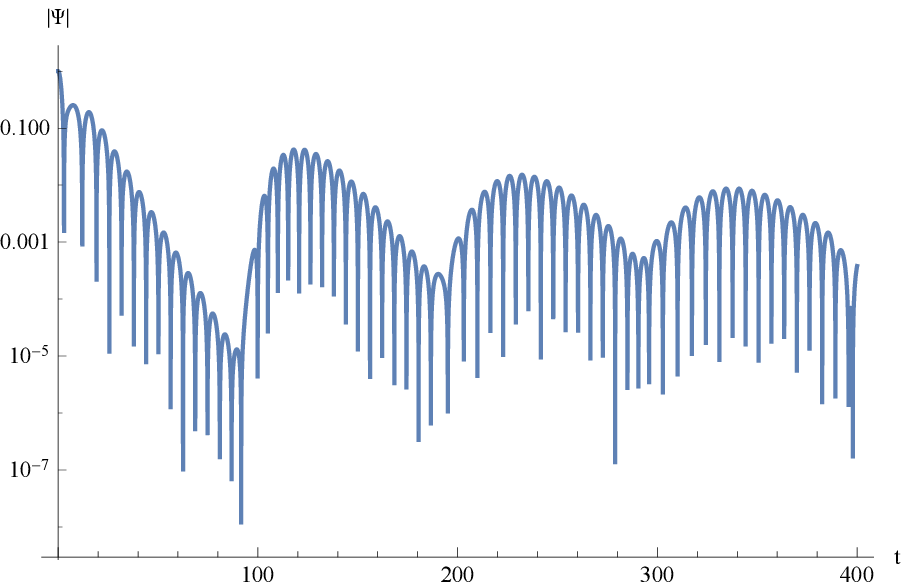}\includegraphics*{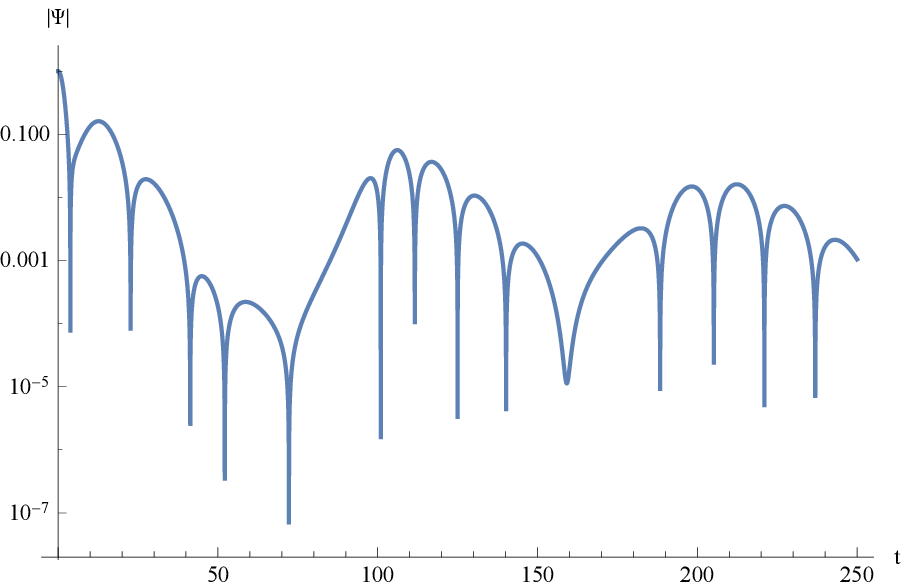}}
\caption{Time-domain profiles for $a=1.01$ for electromagnetic field ($s=1$, $\ell=1$) (left panel) and for scalar field ($s=0$, $\ell=0$) (right panel) in the wormhole spacetime ($M=0.5$). The quasinormal ringing of the initial black-hole like fall-off goes over into the series of echoes. }\label{fig2}
\end{figure*}

\subsection{The time domain method}

If we use the second derivative in time instead of stationary ansatz leading to $\omega^2$ term in Eq. (\ref{wave-equation}), the corresponding partial differential equations can be integrated at a fixed $r$ in the time domain. We use Gaussian initial wavepacket and the technique of integration in the time domain developed by Gundlach, Price and Pullin in \cite{Gundlach:1993tp}, where it was shown that the quasinormal modes do not depend on the parameters of this wavepacket.
 We integrate the wave-like equation rewritten in terms of the light-cone variables $u=t-r_*$ and $v=t+r_*$. The appropriate discretization scheme is:
$$
\Psi\left(N\right)=\Psi\left(W\right)+\Psi\left(E\right)-\Psi\left(S\right)-
$$
\begin{equation}\label{Discretization}
-\Delta^2\frac{V\left(W\right)\Psi\left(W\right)+V\left(E\right)\Psi\left(E\right)}{8}+{\cal O}\left(\Delta^4\right)\,,
\end{equation}
where the following designations for the points were used:
$N=\left(u+\Delta,v+\Delta\right)$, $W=\left(u+\Delta,v\right)$, $E=\left(u,v+\Delta\right)$ and $S=\left(u,v\right)$. The initial data are given on the null surfaces $u=u_0$ and $v=v_0$.
To extract the values of the quasinormal modes out of the obtained time-domain profile, we use the Prony method of fitting the signal by a sum of damped exponents (see, e.g., \cite{Berti:2007dg})
$$
\Psi\left(t\right)\simeq\sum\limits_{i=1}^p C_ie^{-i\omega_it}.
$$
Due to the fact that the signal can be very well approximated by its fundamental mode only and contribution of higher overtones is usually negligible, we extracted the fundamental mode from the profile, although the time-domain profile takes into consideration contributions of all overtones.

By changing the parameters of the numerical integration (the integration grid and the precision of the whole procedure) we can control the error of the calculations: if diminishing of the integration grid and increasing of the precision do not change the time-domain profile, it is a sign of a good accuracy of obtained results. The values of the integration step and precision depend on the parameters of the metric and quantum numbers of the considered fields (such as $\ell$, $n$, $s$). The corresponding code for the numerical integration is written in \emph{Mathematica}\circledR.

\begin{figure*}
\resizebox{\linewidth}{!}{\includegraphics*{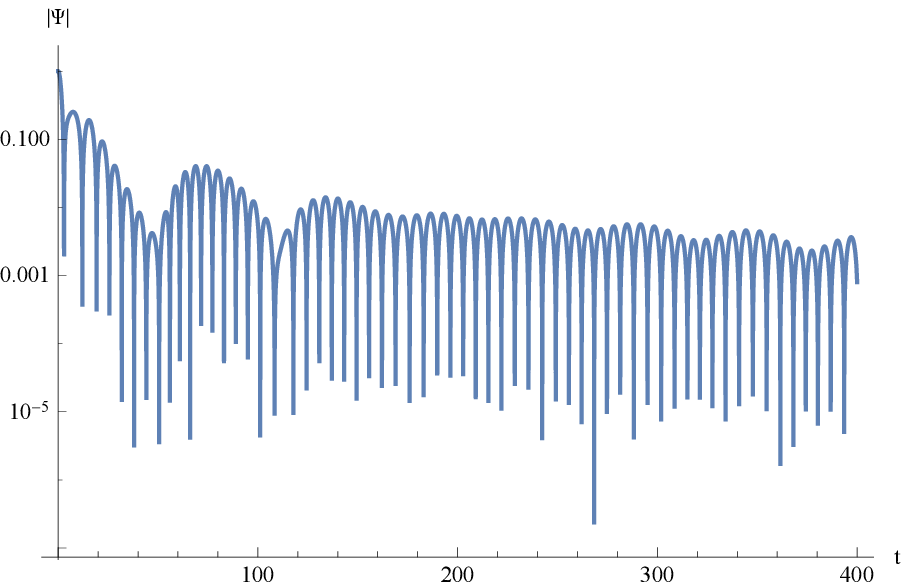}\includegraphics*{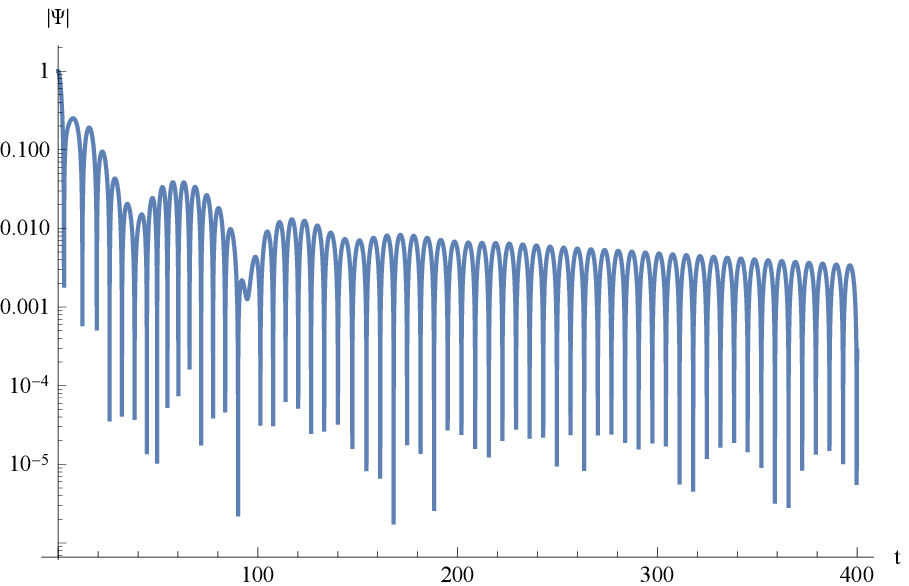}}
\resizebox{\linewidth}{!}{\includegraphics*{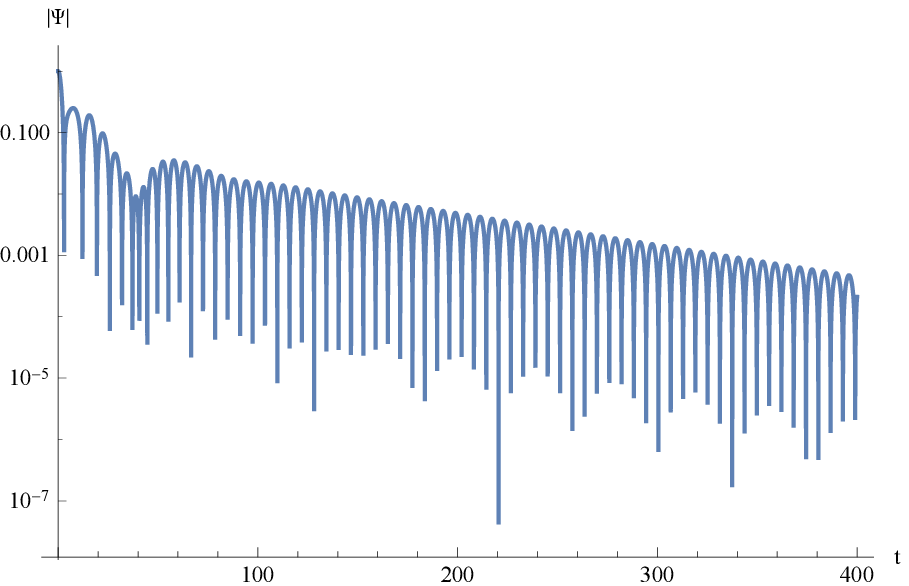}\includegraphics*{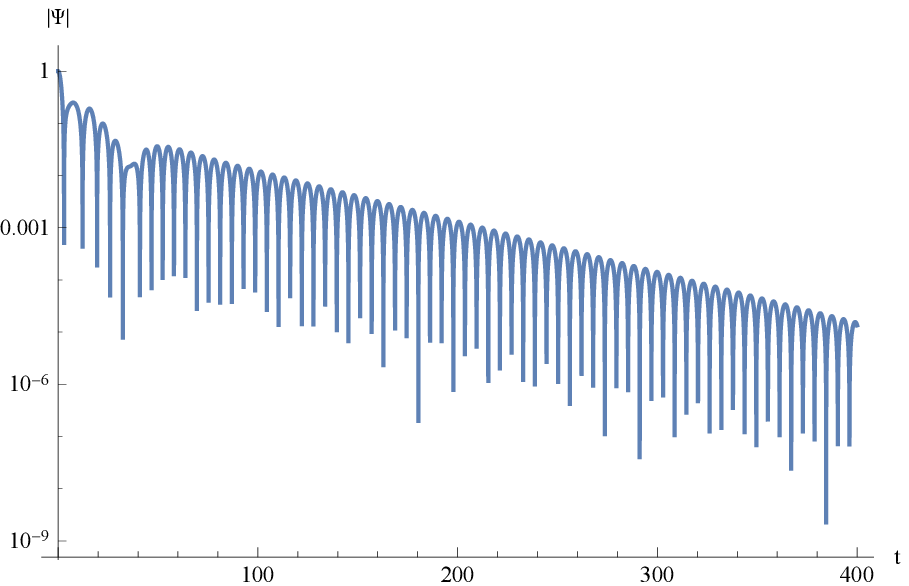}}
\caption{Time-domain profiles in wormhole cases $a=1.05$, $a=1.10$, $a=1.15$, $a=1.20$ (from left to right and up to down, with $M=0.5$) for the electromagnetic field ($s=1$) and $\ell=1$. While the near transition $a=1.05$ case is characterized by echoes, the larger $a$ have a clear quasinormal mode of the wormhole at late times. }\label{fig1}
\end{figure*}

\section{Quasinormal modes}

For the calculation of the quasinormal modes of the continuously parametrized spacetime (\ref{metric}) we fix the mass of the compact object, $M=0.5$, and study the scalar and electromagnetic perturbations for various multipole numbers $\ell$.

In the case of a black hole ($a<1$) or a one-way wormhole ($a=1$), the effective potential has the form of a potential barrier with a single maximum and two turning points (see Fig. \ref{pot0}) so that we can use both the WKB and time-domain methods. Tables 1 and 2 demonstrate a good agreement between the results obtained by these methods. We rely upon the results obtained by the time-domain integration for the case of a traversable wormhole ($a>1$), because the WKB formula (\ref{WKBformula-spherical}) cannot be applied in the case of more than two turning points (see Fig. \ref{pot1}).

For the values of the parameter $a$ in vicinity of the black hole/wormhole transition, the effective potential (Fig. \ref{pot1}) demonstrates two relatively distant peaks. This phenomenon causes modification of the signal at the late times known as echoes, which can occur either because of the modification of the black-hole metric near its event horizon \cite{Cardoso:2016rao,Cardoso:2016oxy} or due to the matter situated at some distance from the compact object \cite{Konoplya:2018yrp,Barausse:2014tra}. Because of the existence of the second peak of the effective potential, we cannot apply the WKB formula and have to use the time-domain integration to determine the quasinormal modes. After the black hole/wormhole transition, we observe the quasinormal ringing of the initial black hole ($t<100$ for electromagnetic field and $t<50$ for scalar field) followed by the series of echoes (see the time-domain profiles on Fig. \ref{fig2} ). From the time-domain profiles on Fig. \ref{fig1} we extracted (for $t \approx 17-37$) the quasinormal mode of the black hole which remained in the form of the initial outburst. Then we observe the echo effect and for $t \approx 200-300$ we obtain the quasinormal mode of the newly formed wormhole. While the parameter $a$ is increasing, both peaks of the effective potential become closer and the echo effect decreases rather quickly, and a distinctive quasinormal mode of the wormhole remains. For $a=1.2$ we have already a kind of the initial outburst (which still slightly resembles the regular black hole quasinormal ringing) and the well seen quasinormal mode of the wormhole spacetime. However, the increasing parameter $a$ implies an earlier start of the numerical noise -- for this reason we have to integrate with smaller grid and higher precision for large values of $a$.

The results are presented in Table 2 -- for values of the parameter $a$ corresponding to the states \textit{after} the black hole/wormhole transition ($a>1$), we list two modes obtained by the time-domain integration. The first mode is the remnant of the black-hole ringing in the initial outburst, and the second mode is the echoe followed by the ringdown of the wormhole. Recall that for these values of the parameter $a$, the WKB formula cannot be applied. Therefore, there are no corresponding results in Table 2.

\begin{center}
\begin{table*}
\begin{tabular}{p{2cm}cc}
\multicolumn{3}{c}{{\bf Table 1.} Fundamental quasinormal mode of the scalar field, $\ell=2$, obtained by the WKB and time-domain methods.} \\
\hline
$a$ & WKB & Time domain  \\[3pt]  \hline \\[-5pt]
0 & $0.967303-0.193537i$ & $0.9673 - 0.1935i$  \\[5pt]
0.1 & $0.967298-0.193099i$ & $0.9673 - 0.1931i$  \\[5pt]
0.2 & $0.967281-0.191780i$ & $0.9673 - 0.1918i$  \\[5pt]
0.3 & $0.967240-0.189565i$ & $0.9672 - 0.1896i$ \\[5pt]
0.4 & $0.967157-0.186428i$ & $0.9671- 0.1865i$ \\[5pt]
0.5 & $0.967002-0.182331i$ & $0.9669 - 0.1824i$ \\[5pt]
0.6 & $0.966729-0.177217i$ & $0.9667 - 0.1773i$ \\[5pt]
0.7 & $0.966270-0.171009i$ & $0.9662- 0.1711i$ \\[5pt]
0.8 & $0.965514-0.163606i$ & $0.9654 - 0.1637i$ \\[5pt]
0.9 & $0.964283-0.154880i$ & $0.9642 - 0.155i$ \\[5pt]
1 & $0.962281-0.144713i$ & $0.9622 - 0.1448i$ \\[5pt]
\hline
\end{tabular}
\end{table*}
\end{center}

\begin{center}
\begin{table*}
\begin{tabular}{p{2cm}cc}
\multicolumn{3}{c}{{\bf Table 2.} Fundamental quasinormal mode of the electromagnetic field, $\ell=1$, obtained by the WKB and time-domain methods.} \\
\hline
$a$ & WKB & Time domain  \\[3pt]  \hline \\[-5pt]
0 & $0.4965-0.1853i$ & $0.4965 - 0.185i$  \\[5pt]
0.1 & $0.4968-0.1849i$ & $0.4968 - 0.1846i$  \\[5pt]
0.2 & $0.4977-0.1837i$ & $0.4975 - 0.1834i$  \\[5pt]
0.3 & $0.4989-0.1815i$ & $0.4986 - 0.1813i$ \\[5pt]
0.4 & $0.50043-0.1784i$ & $0.5002- 0.1784i$ \\[5pt]
0.5 & $0.50234-0.17445i$ & $0.5022 - 0.1744i$ \\[5pt]
0.6 & $0.50455-0.16933i$ & $0.5045 - 0.1693i$ \\[5pt]
0.7 & $0.50691-0.16291i$ & $0.5069- 0.163i$ \\[5pt]
0.8 & $0.50917-0.15500i$ & $0.5092 - 0.1551i$ \\[5pt]
0.9 & $0.51092-0.14544i$ & $0.5109 - 0.1455i$ \\[5pt]
1 & $0.51142-0.13429i$ & $0.5114 - 0.1343i$ \\[5pt]
1.01 &  & $0.5113 - 0.1327i$,\, echo  \\[5pt]
1.05 &  & $0.5167 - 0.1303i$,\, echo  \\[5pt]
1.1 & & $0.512 - 0.1277i$,\, $0.4604-0.0037i$ \\[5pt]
1.15 & & $0.5749-0.0044i$ ,\, $0.5103 - 0.0119i$ \\[5pt]
1.2 & & $0.7137-0.0042i$ ,\, $0.5385 - 0.0223i$ \\[5pt]
1.25 & & initial outburst,\, $0.5549-0.033i$  \\[5pt]
1.3 & & initial outburst,\, $0.5639-0.043i$ \\[5pt]
1.35 & & initial outburst ,\, $0.568 - 0.052i$ \\[5pt]
1.4 & & initial outburst ,\, $0.5686 - 0.0599i$ \\[5pt]
\hline
\end{tabular}
\end{table*}
\end{center}

The results presented in Tables 1 and 2 can be summarized in the following way. We observe three qualitatively different types of quasinormal ringing behaviour:
\begin{itemize}
\item usual damping of the perturbation of the regular black hole;
\item remnants of the initial black-hole ringing, modified by the series of echoes just after the black hole/wormhole transition;
\item distinctive wormhole's mode preceded by an initial outburst, which resembles the decaying profile of the regular black hole.
\end{itemize}

The low multipole numbers $\ell$ are of high interest because of their dominant role in the signal. For large multipole numbers, all the fields usually behave similarly, due to the effective potential being independent of the field spin. Of special interest is the limit of $\ell\rightarrow\infty$, corresponding to the so called eikonal regime, or the regime of geometrical optics, when the behavior of the quasinormal modes is governed by the circular null geodesics \cite{Cardoso:2008bp}; for exceptions of this rule see \cite{Konoplya:2017wot,Toshmatov:2018tyo,Konoplya:2019hml}.

In the eikonal regime, it is possible to obtain even analytical formulae for the quasinormal modes (see \cite{MashhoonBlome}, \cite{Churilova:2019jqx} and references therein). Using the WKB approximation \cite{Schutz:1985zz}:
$$
\omega^2=V_0+\sqrt{-2V_2}\left(n+\frac{1}{2}\right)i,
$$
where $n$ is the overtone number, we expand the result for $\omega$ into the powers of the inverse multipole number ${\ell}^{-1}$ and obtain the following analytical formula for the eikonal quasinormal modes
$$
\omega=\frac{1}{3\sqrt{3}M}\left(\ell+\frac{1}{2}\right)-
$$
\begin{equation}\label{QNM}
-i\frac{1}{3\sqrt{3}M}\left(n+\frac{1}{2}\right)\sqrt{1-\left(\frac{a}{3 M}\right)^2}+{\cal O}\left(\frac{1}{\ell +\frac{1}{2}}\right),
\end{equation}
where $M$ is the mass. The eikonal formula given above is valid whenever the effective potential has the form of the potential barrier with a single maximum, implying existence of two turning points, as is the black hole case $a<2M$.

Taking, for an illustration, $\ell=100$, we obtain from the formula (\ref{QNM}) for the fundamental mode (with $M=0.5$, $a=0.5<2M$) $\omega=38.6825 - 0.181444i$, and using the WKB method with Pad\'{e} approximation, we obtain $\omega=38.68138-0.18144i$, demonstrating a good concordance between the analytical results and numerical results.

\section{Conclusions}

For the spacetime model (\ref{metric}) interpolating between a regular black hole and a wormhole we studied the ringing of the black hole/wormhole transition. We showed that this transition is characterized by echoes and observed three qualitatively distinct types of quasinormal ringing behaviour:
\begin{itemize}
\item usual ringing of the regular black hole;
\item remnant of the initial black-hole ringing, followed by the series of echoes directly after the black hole/wormhole transition;
\item clear wormhole's mode, before which comes an initial outburst, similar to the profile of the regular black hole.
\end{itemize}

 The results obtained by the WKB method with Pad\'{e} approximants, and by the time-domain integration, are in good concordance in the common range of their applicability. For the quasinormal modes in the eikonal regime we found analytical formula.

It would be interesting to study also massive fields in the vicinity of the transition between the regular black hole and wormhole states as arbitrarily long lived quasinormal modes which exist not only for black holes (see for example \cite{Ohashi:2004wr,Konoplya:2017tvu,Zinhailo:2018ska,Churilova:2019sah}), but also for wormholes with non-zero tidal force in the radial direction \cite{Churilova:2019qph} could interfere with echoes at late times. Our approach could also be extended to the case of charged wormholes suggested recently in \cite{Huang:2019arj}.

\acknowledgments{The authors acknowledge  the  support  of  the  grant  19-03950S of Czech Science Foundation ($GA\check{C}R$). M. S. C. acknowledges hospitality and support of Silesian University in Opava and Roman Konoplya for useful discussions. The authors also would like to thank anonymous referees for valuable advice.}

\end{document}